\documentclass[secnumarabic,amssymb, nobibnotes, aps, prd]{revtex4-1}
\usepackage{amsmath} \usepackage{amssymb}
\usepackage{graphicx} 
\usepackage{epstopdf}
\usepackage{amsmath}
\usepackage{float}
\usepackage{amsmath,amssymb,amsthm,amsfonts,mathrsfs,bm,verbatim}
\usepackage{graphicx,subfigure}

\newcommand{\bea}{\begin{eqnarray}}
\newcommand{\eea}{\end{eqnarray}}
\def\nn{\nonumber}

\newcommand{\bgn}{\begin{align}}
\newcommand{\egn}{\end{align}}

\def\o{\omega} \def\l{\lambda}

\allowdisplaybreaks[1]

\begin{document}

\title{Six-dimensional non-extremal Reissner-Nordstrom black hole, charged massive scalar perturbation and black hole bomb}

\author{Run-Dong Zhao}
\affiliation{Guangdong Provincial Key Laboratory of Nuclear Science, Institute of quantum matter,South China Normal University, Guangzhou 510006, China}
\affiliation{Guangdong-Hong Kong Joint Laboratory of Quantum Matter, Southern Nuclear Science Computing Center, South China Normal University, Guangzhou 510006, China}

\author{Jia-Hui Huang}
\email{huangjh@m.scnu.edu.cn}
\affiliation{Guangdong Provincial Key Laboratory of Quantum Engineering and Quantum Materials,
School of Physics and Telecommunication Engineering,
South China Normal University, Guangzhou 510006, China}
\affiliation{Institute of quantum matter,South China Normal University, Guangzhou 510006, China}
\affiliation{Guangdong-Hong Kong Joint Laboratory of Quantum Matter, Southern Nuclear Science Computing Center, South China Normal University, Guangzhou 510006, China}

\begin{abstract}
The superradiant stability of higher dimensional non-extremal Reissner-Nordstrom black hole under charged massive scalar perturbation is analytically studied. We extend our previous studies of four- and five-dimensional non-extremal Reissner-Nordstrom black hole cases to six-dimensional case. By analyzing the derivative of the effective potential with an analytical method, we find that no potential well exists outside the outer horizon of the black hole for the superradiant scalar modes. This means
that there is no black hole bomb for the system consisting of six-dimensional Reissner-Nordstrom black hole and charged massive scalar perturbation and the system is superradiantly stable.
\end{abstract}

\maketitle
\tableofcontents

\section{Introduction}
Einstein's general relativity is the current description of the gravitational force. One important prediction of general relativity is black hole.
In early 1970's, it was pointed that rotating energy might be extracted from a classical rotating black hole by observers outside the black hole  \cite{Penrose:1971uk,Bardeen:1972fi,Press:1972zz,Bekenstein:1973mi}.
When a charged bosonic wave is imping on a charged rotating black hole, the scattered wave may be amplified by the black hole if the wave frequency $\omega$ obeys
  \begin{equation}\label{superRe}
   0<\omega < \text{m}\Omega_H  + e\Phi_H,
  \end{equation}
where $e$ and $\text{m}$ are respectively the charge and azimuthal number of the bosonic wave, $\Omega_H$ is the angular velocity of the black hole horizon and $\Phi_H$ is the electromagnetic potential of the black hole horizon. This wave amplification process is called superradiance, which appears in various areas of physics(for a recent review, see\cite{Brito:2015oca}).

When there is a mirror-like mechanism that makes the amplified wave be scattered back and forth between the mirror and the black hole, the background black hole geometry will become superradiantly unstable. This is dubbed the black hole bomb mechanism \cite{Press:1972zz,Cardoso:2004nk,Herdeiro:2013pia,Degollado:2013bha}. For the massive bosonic perturbation, its  mass term behaves as a natural mirror for the low energy perturbation modes. The superradiant (in)stability of asymptotically flat rotating black holes under massive scalar and vector perturbation has been studied extensively in the literature \cite{Strafuss:2004qc,Konoplya:2006br,Cardoso:2011xi,Dolan:2012yt,Hod:2012zza,Hod:2014pza,Aliev:2014aba,Hod:2016iri,Degollado:2018ypf,Lin:2021ssw,Xu:2020fgq,Huang:2019xbu,Ponglertsakul:2020ufm,East:2017ovw,East:2017mrj}.

Four-dimensional asymptotically flat charged Reissner-Nordstrom (RN) black holes were proved superradiantly stable against charged massive scalar perturbation \cite{Hod:2013eea,Huang:2015jza,Hod:2015hza,DiMenza:2014vpa,Chowdhury:2019ptb,Mai:2021yny,Richarte:2021fbi}. The effect of a RN black hole geometry on the motion of the scalar perturbation can be described by an effective potential outside the black hole horizon. Two necessary conditions for the black hole to be superradiantly unstable are the existence of superradiant scalar perturbation modes and the existence of a trapping potential well for the effective potential outside the black hole horizon \cite{Li:2014gfg,Sanchis-Gual:2015lje,Fierro:2017fky,Li:2014fna,Li:2015mqa}. When these two conditions can not be satisfied simultaneously, the black hole and scalar perturbation system will be superradiantly stable. Higher dimensional RN black holes have also been studied numerically in literature \cite{Konoplya:2011qq,Konoplya:2007jv,Konoplya:2008au,Konoplya:2013sba,Konoplya:2008rq,Kodama:2003kk,Kodama:2007sf,Ishibashi:2011ws,Ishihara:2008re,Ishibashi:2003ap}.
Asymptotically flat RN black holes in $D=5,6,..,11$ were shown to be stable under massless scalar perturbation with a numerical method \cite{Konoplya:2008au}.

Recently, we have developed an analytical method in the study of superradiant stability of higher dimensional extremal and non-extremal RN black holes \cite{Huang:2021dpa,Huang:2021jaz,Huang:2022nzm}. It is found that five- and six-dimensional extremal RN black holes are superradiantly stable under charged massive scalar perturbation\cite{Huang:2021dpa}. Even for the $D$-dimensional extremal RN black hole, it is found that there is no black hole bomb for the black hole and scalar perturbation system\cite{Huang:2022nzm}.
It was also proved that there is also no black hole bomb for five-dimensional non-extremal RN black hole under charged massive scalar perturbation and the system is superradiantly stable\cite{Huang:2021jaz}.

In this paper, we extend our previous studies to the six-dimensional non-extremal RN black hole case. We will prove that there is no black hole bomb for six-dimensional non-extremal RN black hole and charged massive scalar perturbation system.
The organization of this paper is as follows. In Section 2, we give a description of the motion of scalar perturbation in six-dimensional RN black hole background.
In Section 3, we analyze the effective potential for the superradiant modes and present the analytical proof that there is no potential well for the effective potential.
There is no black hole bomb for the superradiant modes and the system is superradiantly stable.
The last Section is devoted to a summary.

\section{Scalar Perturbation in $D$=6 RN black hole}
The metric of $D=6$ RN black hole is
\begin{align}
    ds_{6}^2=-f(r)dt^2+\frac{dr^2}{f(r)}+r^2d\Omega_{4}^2,
\end{align}
where
\begin{align}
    f(r)=1-\frac{2m}{r^3}+\frac{q^2}{r^6}.
\end{align}
The parameters $m$ and $q$ are respectively related with the ADM mass $M$ and the electric charge $Q$ of the black hole,
\begin{align}
    m=\frac{3}{4\pi}M, q=\frac{3}{2\sqrt{6}\pi}Q.
\end{align}
The inner and outer horizons of the black hole are
\begin{align}
    r_{\pm}=(m\pm\sqrt{m^2-q^2})^{1/3}.
\end{align}
The components of the background electromagnetic potential is
\begin{align}
    A_{\nu}=(-\frac{2q}{\sqrt{6}r^3},0,0,0,0,0)=(-\frac{c_6 q}{r^3},0,0,0,0,0).
\end{align}

The dynamics of a minimally coupled charged massive scalar perturbation field $\Psi$ with mass $\mu$ and charge $e$ is governed by the Klein-Gordon equation
\begin{align}
    (D_{\nu}D^{\nu}-\mu^2)\Psi=0,
\end{align}
where $D_{\nu}=\nabla_{\nu}-ieA_{\nu}$ is the covariant derivative.
The above equation of motion can be separated into radial part and angular part\cite{Huang:2021jaz,Huang:2021dpa}. The solution of the above equation with definite angular frequency can be decomposed as $\Psi=\sum\limits_{lm} R_{lm}(r) Y_{lm}(\theta_i,\phi) e^{-i\omega t}$. The eigenfunction $Y_{lm}(\theta_i,\phi)$ of the angular equation of motion is scalar harmonics on $S^4$ with eigenvalue $\lambda=l(l+3),(l\geq 0)$. The radial equation of motion is
\begin{align}
    \Delta\frac{d}{dr}(\Delta\frac{dR}{dr})+UR=0,\label{radial_equation}
\end{align}
where $R$ is the radial function (we neglect the indices $l,m$ for simplicity), $\Delta=r^4f(r)$, and
\begin{align}
    U=\left(\omega-c_6\frac{eq}{r^3}\right)^2r^8-l(l+3)r^2\Delta-\mu^2r^4\Delta.
\end{align}
In order to study the superradiant stability of the RN black hole under the massive charged perturbation, the asymptotic solutions of the radial equation near the horizon and at spatial infinity should be considered with appropriate boundary conditions. By defining the tortoise coordinate $y$ by $dy=\frac{r^4}{\Delta}dr$ and a new radial function $\tilde{R}=r^2R$, the radial equation \eqref{radial_equation} can be rewritten as
\begin{align}
    \frac{d^2\tilde{R}}{dy^2}+\tilde{U}\tilde{R}=0,
\end{align}
where
\begin{align}
    \tilde{U}=\frac{U}{r^8}-\frac{4f(r)[2f(r)+2rf'(r)]}{4r^2}.
\end{align}
The asymptotic behaviors of $\tilde{U}$ at the spatial infinity and outer horizon are
\begin{align}
    \lim_{r\rightarrow +\infty}\tilde{U}=\omega^2-\mu^2, ~~\lim_{r\rightarrow r_{+}}\tilde{U}=(\omega-c_6\frac{eq}{r_{+}^3})^2=(\omega-e\Phi_{H})^2,
\end{align}
where $\Phi_{H}$ is the electric potential of the RN black hole at the outer horizon. The chosen boundary conditions are ingoing wave at the outer horizon $(y\rightarrow-\infty)$ and bound states (exponentially decaying modes) at spatial infinity $(y\rightarrow+\infty)$. Then the radial wave equation has the following asymptotic solutions
\bea
&&y\rightarrow+\infty (r\rightarrow+\infty), ~~ \tilde{R}\sim  e^{-\sqrt{\mu^2-\omega^2}y},\\
 &&y\rightarrow-\infty (r\rightarrow r_{+}),  ~~\tilde{R}\sim e^{-i(\omega-e\Phi_{H})y}.
\eea
It is obvious to see that bound state condition at spatial infinity requires the following inequality
\begin{align}
    \omega^2<\mu^2.
\end{align}
The superradiance condition in this case is
\begin{align}
    0<\omega<\omega_{c}=e\Phi_{H}=\frac{c_6 eq}{r_{+}^3}.
\end{align}

\section{Analysis of the effective potential}
We define a new radial function $\psi=\Delta^{1/2}R$, then the radial equation \eqref{radial_equation} can be rewritten as a Schrodinger-like equation
\begin{align}
    \frac{d^2\psi}{dr^2}+(\omega^2-V)\psi=0,
\end{align}
where the effective potential $V$ is
\begin{align}
    V=\omega^2+\frac{B}{A},
\end{align}
and
\begin{align}
    A&=4 r^2 \left(-2 m r^3+q^2+r^6\right)^2,\\
    B&=4 r^{14} \left(\mu ^2-\omega ^2\right)+(4\lambda+8)r^{12}-8 r^{11} \left(\mu ^2 m-c_6 e q \omega \right)-4 (2 \lambda +8) m r^9+4 q^2 r^8 \left(\mu ^2-c_6^2 e^2\right)\nn\\
    &-2 r^6 \left(2 m^2-(2 \lambda +26) q^2\right)-32 m q^2 r^3+8 q^4.
\end{align}
The asymptotic behaviors of the effective potential $V$ near the outer horizon and at spatial infinity are
\bea
    V(r\rightarrow r_{+})&\rightarrow-\infty,~~
     V(r\rightarrow +\infty)&\rightarrow\mu^2+\frac{\lambda+2}{r^2}+o(\frac{1}{r^3}).
\eea
The asymptotic behavior of the derivative of $V$ at spatial infinity is
\begin{align}
    V'=-\frac{2(\lambda+2)}{r^3}+o(\frac{1}{r^4}).
\end{align}
Based on the above asymptotic behaviors, we can know that there is at least one maximum for $V$ outside the outer horizon $r_{+}$ and
there is no potential well near spatial infinity.

In the next part of this section, we will show that there is in fact only one maximum outside the outer horizon for $V$ and no trapping potential well exists by analyzing the real roots of $V'(r)$ outside the horizon $r_+$.
The explicit expression of the derivative of the effective potential $V$ is
\bea
V'(r)=-\frac{2C(r)}{r^3 (r^6-2mr^3 +q^2)^3},
\eea
where
\bea
C(r)&=&(\lambda +2) r^{18}+r^{17} \left(3c_6 e q \omega +3 \mu ^2 m-6 m \omega ^2\right)+r^{15} (-\lambda  m-12 m)\nn\\
&+&r^{14} \left(-6 m^2 \mu^2 - 3 c_6^2 e^2 q^2 - 3 \mu^2 q^2 + 6 c_6 e m q \o + 6 q^2 \o^2\right)+r^{12} \left(-2 \lambda  m^2-12 m^2-\lambda  q^2+42 q^2\right)\nn\\
&+&r^{11} \left(9 \mu^2 m q^2-9c_6 e q^3 \omega \right)+r^9 \left(2 m^3+5 \lambda  m q^2-42 m q^2\right)+r^8 \left(3c_6^2 e^2 q^4-3 \mu ^2 q^4\right)\nn\\
&+&r^6 \left(42 m^2 q^2-2 \lambda  q^4-12 q^4\right)-12 m q^4 r^3+2q^6.
\eea
Defining a new variable $z=r-r_+$ and using relations $m=\frac{1}{2} \left(r_-^3+r_+^3\right)$ and $q^2=r_+^3r_-^3$, we can rewrite $C(r)$ as
\bea
C(r)&=&\tilde{C}(z)=b_{18}z^{18}+b_{17}z^{17}+b_{16}z^{16}+...+b_{2}z^{2}+b_{1}z+b_0,\label{derivative of potential}
\eea
where
\begin{align}
b_{18}&=\lambda +2,~
b_{17}=3(m \mu^2 + c_6 e q \o - 2m \o^2) + 18 r_+ (2 + \lambda),\\
b_{16}&=51 (m \mu^2 r_+ + c_6 e q r_+ \o - 2 m r_+ \o^2) + 153 r_+^2 (2 + \lambda),\\
b_{15}&=-12 m + 408 m \mu^2 r_+^2 + 408 c_6 e q r_+^2 \o - 816 m r_+^2 \o^2 - m \lambda + 816 r_+^3 (2 + \lambda),\\
b_{14}&=-3 c_6^2 e^2 q^2 + 90 (67 r_+^4 - r_+ r_-^3) + 3 \mu^2 (679 r_+^6 + 676 r_+^3 r_-^3 - r_-^6)/2 +
 3 c_6 e q (681 r_+^3 + r_-^3) \o \nn\\&- 6 (340 r_+^6 + 339 r_+^3 r_-^3) \o^2 +
 15 (407 r_+^4 - r_+ r_-^3) \l/2,\\
b_{13}&=-42 c_6^2 e^2 q^2 r_+ + 126 (131 r_+^5 - 5 r_+^2 r_-^3) + 21 \mu^2 (169 r_+^7 + 166 r_+^4 r_-^3 - r_+ r_-^6) +
 42 c_6 e q r_+ (171 r_+^3 + r_-^3) \o\nn\\& - 84 (85 r_+^7 + 84 r_+^4 r_-^3) \o^2 + 21 (811 r_+^5 - 5 r_+^2 r_-^3)\l/2,\\
b_{12}&=-273 c_6^2 e^2 q^2 r_+^2 + 34395 r_+^6 - 2694 r_+^3 r_-^3 - 3 r_-^6 + 273 \mu^2 (67 r_+^8 + 64 r_+^5 r_-^3 - r_+^2 r_-^6)/2 \nn\\&+
 273 c_6 e q r_+^2 (69 r_+^3 + r_-^3) \o - 546 (34 r_+^8 + 33 r_+^5 r_-^3) \o^2 + (36672 r_+^6 - 459 r_+^3 r_-^3 - r_-^6) \l/2,\\
b_{11}&=-1092 c_6^2 e^2 q^2 r_+^3 + 18 (3079 r_+^7 - 431 r_+^4 r_-^3 - 2 r_+ r_-^6) + 3 \mu^2 (12012 r_+^9 + 10923 r_+^6 r_-^3 - 361 r_+^3 r_-^6)/2 \nn\\&+ c_6 e q (38220 r_+^6 + 1083 r_+^3 r_-^3) \o - 2184 (17 r_+^9 + 16 r_+^6 r_-^3) \o^2 +
 3(20757 r_+^7 - 471 r_+^4 r_-^3 - 4 r_+ r_-^6) \l/2,\\
b_{10}&=-3003 c_6^2 e^2 q^2 r_+^4 + 33\mu^2 r_+^4 (1677 r_+^6 + 1407 r_+^3 r_-^3 - 88 r_-^6)/2 +
 198 (350 r_+^8 - 79 r_+^5 r_-^3 - r_+^2 r_-^6) \nn\\&+ 363 c_6 e q (169 r_+^7 + 8 r_+^4 r_-^3) \o - 858 r_+^7 (68 r_+^3 + 61 r_-^3) \o^2 +
 33(2559 r_+^8 - 99 r_+^5 r_-^3 - 2 r_+^2 r_-^6) \l/2,\\
b_{9}&=-6006 c_6^2 e^2 q^2 r_+^5 + \frac{33}{2} \mu^2 r_+^5 (2028 r_+^6 + 1497 r_+^3 r_-^3 - 167 r_-^6) +
 (266201 r_+^9 - 88521 r_+^6 r_-^3 - 2721 r_+^3 r_-^6 + r_-^9)/4\nn\\& + c_6 e q (78936 r_+^8 + 5511 r_+^5 r_-^3) \o -
 858 r_+^8 (85 r_+^3 + 71 r_-^3) \o^2 + 5(18403 r_+^9 - 1176 r_+^6 r_-^3 - 43 r_+^3 r_-^6) \l/2,\\
b_{8}&=3 c_6^2 e^2 q^4 - 9009 c_6^2 e^2 q^2 r_+^6 + 3\mu^2 r_+^6 (21307 r_+^6 + 12793 r_+^3 r_-^3 - 2510 r_-^6)/2 +
 9 r_+^4 (4477 r_+^6 - 465 r_+^3 r_-^3 - 25 r_-^6) \l\nn\\& + c_6 e q (81939 r_+^9 + 7524 r_+^6 r_-^3) \o -
 858 r_+^9 (85 r_+^3 + 64 r_-^3) \o^2 + 9r_+ (21077 r_+^9 - 9321 r_+^6 r_-^3 - 741 r_+^3 r_-^6 + r_-^9)/4,\\
b_{7}&=24 c_6^2 e^2 q^4 r_+ - 10296 c_6^2 e^2 q^2 r_+^7 + 3 \mu^2 r_+^7 (8008 r_+^6 + 3355 r_+^3 r_-^3 - 1229 r_-^6) +
 \frac{9}{2} r_+^5 (6269 r_+^6 - 1047 r_+^3 r_-^3 - 68 r_-^6) \l \nn\\&+ 66 c_6 e q r_+^7 (1040 r_+^3 + 111 r_-^3) \o -
 3432 r_+^{10} (17 r_+^3 + 11 r_-^3) \o^2 +9 r_+^2 (2519 r_+^9 - 1203 r_+^6 r_-^3 - 345 r_+^3 r_-^6 + r_-^9) ,\\
b_{6}&=84 c_6^2 e^2 q^4 r_+^2 - 9009 c_6^2 e^2 q^2 r_+^8 + \frac{21}{2} \mu^2 r_+^8 (1339 r_+^6 + 250 r_+^3 r_-^3 - 239 r_-^6) +
\frac{1}{2} r_+^6 (31199 r_+^6 - 8281 r_+^3 r_-^3 - 508 r_-^6) \l\nn\\&+ 21 c_6 e q r_+^8 (2197 r_+^3 + 231 r_-^3) \o -
 546 r_+^{11} (68 r_+^3 + 35 r_-^3) \o^2 +  \frac{9}{2} r_+^3 (966 r_+^9 + 343 r_+^6 r_-^3 - 992 r_+^3 r_-^6 + 7 r_-^9) ,\\
b_{5}&=168 c_6^2 e^2 q^4 r_+^3 - 6006 c_6^2 e^2 q^2 r_+^9 + 21 \mu^2 r_+^9 (299 r_+^6 - 31 r_+^3 r_-^3 - 52 r_-^6) + \frac{3}{2} r_+^7 (4447 r_+^6 - 1847 r_+^3 r_-^3 - 62 r_-^6) \l\nn\\& + 42 c_6 e q r_+^9 (585 r_+^3 + 44 r_-^3) \o -
 1092 r_+^{12} (17 r_+^3 + 6 r_-^3) \o^2 - \frac{27}{2} r_+^4 (239 r_+^9 - 593 r_+^6 r_-^3 + 361 r_+^3 r_-^6 - 7 r_-^9),\\
b_{4}&=210 c_6^2 e^2 q^4 r_+^4 - 3003 c_6^2 e^2 q^2 r_+^{10} - \frac{27}{2} r_+^5 (261 r_+^3 - 14 r_-^3) (r_+^3 - r_-^3)^2 +
 \frac{3}{2} \mu^2 r_+^{10} (1379 r_+^6 - 634 r_+^3 r_-^3 - 151 r_-^6)\nn\\& + 3 c_6 e q r_+^{10} (3381 r_+^3 + 11 r_-^3) \o -
 42 r_+^{13} (170 r_+^3 + 27 r_-^3) \o^2 + \frac{15}{2} r_+^8 (284 r_+^6 - 181 r_+^3 r_-^3 + 5 r_-^6) \l,\\
b_{3}&=168 c_6^2 e^2 q^4 r_+^5 - 1092 c_6^2 e^2 q^2 r_+^{11} - 9 r_+^6 (193 r_+^3 - 25 r_-^3) (r_+^3 - r_-^3)^2 +
 \frac{3}{2} \mu^2 r_+^{11} (316 r_+^6 - 281 r_+^3 r_-^3 + 19 r_-^6) \nn\\&+ c_6 e q (3132 r_+^{14} - 393 r_+^{11} r_-^3) \o -
 24 r_+^{14} (85 r_+^3 - 6 r_-^3) \o^2 + \frac{3}{2} r_+^9 (319 r_+^6 - 305 r_+^3 r_-^3 + 40 r_-^6) \l,\\
b_{2}&=84 c_6^2 e^2 q^4 r_+^6 - 273 c_6^2 e^2 q^2 r_+^{12} -\frac{27}{2} r_+^7 (38 r_+^3 - 11 r_-^3) (r_+^3 - r_-^3)^2 +
 \frac{27}{2} \mu^2 r_+^{12} (5 r_+^6 - 7 r_+^3 r_-^3 + 2 r_-^6)\nn\\& + 3 c_6 e q r_+^{12} (227 r_+^3 - 74 r_-^3) \o +
 6 r_+^{15} (-68 r_+^3 + 23 r_-^3) \o^2 +27 r_+^{10} (5 r_+^6 - 7 r_+^3 r_-^3 + 2 r_-^6) \l/2,\\
b_{1}&=24 c_6^2 e^2 q^4 r_+^7 - 42 c_6^2 e^2 q^2 r_+^{13} + 9 \mu^2 r_+^{13} (r_+^3 - r_-^3)^2/2 - 27 r_+^8 (13 r_+^3 - 7 r_-^3) (r_+^3 - r_-^3)^2/4 \nn\\&+ c_6 e q (93 r_+^{16} - 57 r_+^{13} r_-^3) \o + (-51 r_+^{19} +
    33 r_+^{16} r_-^3) \o^2 + \frac{9}{2}r_+^{11} (r_+^3 - r_-^3)^2 \l,\\
b_{0}&=-\frac{3}{4}r_+^9(r_+^3-r_-^3)[9(r_+^3-r_-^3)^2+4 (c_6 e q r_+ - r_+^4 \o)^2].
\end{align}

In the following, we will analyze the signs or sign relations of the coefficients $b_i (i=0,1,2,...,18)$ and prove that there is no potential well for the effective potential $V$ outside the horizon $r_+$. First, it is easy to see that
\begin{align}
    b_{18}>0,~b_0<0.\label{ineq1}
\end{align}
Next, let's see the sign of $b_{17}$. From the superradiance condition $\omega<\frac{c_6 eq}{r_{+}^3}$ and the inequality $m=\frac{1}{2} \left(r_-^3+r_+^3\right)<r_+^3$, we have $m\o<c_6 eq$. Together with the bound state condition $\omega^2 < \mu^2$, we have
\begin{align}\label{ineq2}
b_{17}=3 m \mu^2 + 3 c_6 e q \o - 6 m \o^2 + 18 r_+ (2 + \lambda)=3m(\mu^2-\o^2)+3\o(c_6 e q-m\o)+18 r_+ (2 + \lambda)>0.
\end{align}
Similarly, we can obtain
\bea
b_{16}&=&51 m \mu^2 r_+ + 51 c_6 e q r_+ \o - 102 m r_+ \o^2 + 153 r_+^2 (2 + \lambda)\nn\\\label{ineq3}
&=&51mr_+(\mu^2-\o^2)+51r_+\o(c_6 e q-m\o)+ 153 r_+^2 (2 + \lambda)>0.\\
b_{15}&=&-12 m + 408 m \mu^2 r_+^2 + 408 c_6 e q r_+^2 \o - 816 m r_+^2 \o^2 - m \lambda + 816 r_+^3 (2 + \lambda)\nn\\
&=&408 m r_+^2 (\mu^2-\o^2)+ 408r_+^2 \o(c_6 e q-m\o)+1632r_+^3 -12m+(816 r_+^3-m)\lambda>0. \label{ineq4}
\eea

It is difficult to determine the signs of other coefficients. Next, we explore the sign relations between pairs of adjacent coefficients in the following sequence
\bea
(b_{14},b_{13},b_{12},...,b_3,b_2,b_1).
\eea
We define new scaled coefficients as follows,
\bea
&&b'_i=\frac{b_i}{R_i},(i=1,...,14),~\\
&&R_{14}=\frac{3}{2} (681 r_+^6 + 680 r_+^3 r_-^3 + r_-^6),~R_{13}=21 r_+ (171 r_+^6 + 170 r_+^3 r_-^3 + r_-^6),\nn\\
&&R_{12}=\frac{273}{2} r_+^2 (69 r_+^6 + 68 r_+^3 r_-^3 + r_-^6),~R_{11}=\frac{3}{2} (12740 r_+^9 + 12373 r_+^6 r_-^3 + 361 r_+^3 r_-^6),\nn\\&&R_{10}=\frac{33}{2} r_+^4 (1859 r_+^6 + 1765 r_+^3 r_-^3 + 88 r_-^6),~R_9=\frac{33}{2}r_+^5 (2392 r_+^6 + 2195 r_+^3 r_-^3 + 167 r_-^6),\nn\\&&R_{8}=\frac{3}{2} r_+^6 (27313 r_+^6 + 23815 r_+^3 r_-^3 + 2510 r_-^6),~R_{7}=3 r_+^7 (11440 r_+^6 + 9229 r_+^3 r_-^3 + 1229 r_-^6), \nn\\&&R_{6}=\frac{21}{2} r_+^8 (2197 r_+^6 + 1570 r_+^3 r_-^3 + 239 r_-^6),~R_{5}=21 r_+^9 (585 r_+^6 + 343 r_+^3 r_-^3 + 52 r_-^6), \nn\\&&R_{4}=\frac{3}{2} r_+^{10} (3381 r_+^6 + 1390 r_+^3 r_-^3 + 151 r_-^6),~R_{3}=\frac{3}{2} r_+^{11} (1044 r_+^6 + 185 r_+^3 r_-^3 - 19 r_-^6), \nn\\&&R_{2}=\frac{3}{2} r_+^{12} (227 r_+^6 - 29 r_+^3 r_-^3 - 18 r_-^6),~ R_{1}=\frac{3}{2} r_+^{13} (31 r_+^6 - 16 r_+^3 r_-^3 - 3 r_-^6).
\eea
Let's see the difference between $b'_{14}$ and $b'_{13}$. After a straightforward calculation, we obtain
\bea
b'_{14}-b'_{13}=\frac{63r_+^2}{4 R_{14}R_{13}}[2040 r_+^2 (r_+^3 + r_-^3)^3(\mu^2-\o^2)+2040r_+^2c_6eq(c_6eq(r_+^3+r_-^3)-2
r_+^3r_-^3\o)\nn\\
-5 r_-^9 (12 + \l) + 3 r_+^3 r_-^6 (8956 + 1653 \l) + 5 r_+^6 r_-^3 (63636 + 28423 \l) + 3 r_+^9 (101436 + 47893 \l)].
\eea
Given the bound state condition $\o^2<\mu^2$ and superradiance condition $0<\o r_+^3<c_6 eq$, it is easy to prove that
\bea
b'_{14}-b'_{13}>0.
\eea
The possible signs of $(b'_{14},b'_{13})$ are $(+,+),(+,-),(-,-)$. So we have the sign relation $\text{sign}(b'_{14})\geqslant\text{sign}(b'_{13})$. Because the scaled factors $R_i(i=1,...,14)$ are all positive, then we also have the following sign relation
\bea
\text{sign}(b_{14})\geqslant\text{sign}(b_{13}).\label{sign1413}
\eea
For the difference between $b'_{13}$ and $b'_{12}$, after a straightforward calculation, we obtain
\bea
b'_{13}-b'_{12}=\frac{21r_+}{4 R_{13}R_{12}}[111384 r_+^5 (r_+^3 + r_-^3)^3(\mu^2-\o^2)+111384r_+^5c_6eq(c_6eq(r_+^3+r_-^3)-2
r_+^3r_-^3\o)+2 r_-^{12} (6 + \l) \nn\\- r_+^3 r_-^9 (3564 + 107 \l) + 3 r_+^6 r_-^6 (337236 + 70547 \l) +
 9 r_+^{12} (676176 + 303887 \l) + 3 r_+^9 r_-^3 (2168828 + 883239 \l)].
\eea
Given the bound state condition $\o^2<\mu^2$ and superradiance condition $0<\o r_+^3<c_6 eq$, it is also easy to obtain that
\bea
b'_{13}-b'_{12}>0.
\eea
So we have the following sign relation
\bea
\text{sign}(b_{13})\geqslant\text{sign}(b_{12}).\label{sign1312}
\eea
For the difference between $b'_{12}$ and $b'_{11}$, we obtain
\bea
&&b'_{12}-b'_{11}=\frac{3r_+^3}{4R_{12}R_{11}}\{[1092 r_+^5 (6188 r_+^9 + 18462 r_+^6 r_-^3 + 18363 r_+^3 r_-^6 + 6089 r_-^9)(\mu^2-\o^2)]+731 r_-^{12} (6 + \l)
\nn\\&&+[546 c_6^2 e^2 q^2 r_+^2 (12376 r_+^6 + 12379 r_+^3 r_-^3 + 3 r_-^6) -
 1092 c_6 e q r_+^5 r_-^3 (12376 r_+^3 + 3 r_-^3) \o] + r_+^3 r_-^9 (-161814 + 24767 \l)\nn\\&& +
 273 r_+^{12} (660788 + 279127 \l) + 2 r_+^6 r_-^6 (22234659 + 5349488 \l) + 3 r_+^9 r_-^3 (64672614 + 23811425 \l)\}.
\eea
In the above equation, it is easy to see that the term in the square bracket involving $(\mu^2-\o^2)$ is positive with respect to the bound state condition. The term in the square bracket involving $\o$ is
\bea
&&[546 c_6^2 e^2 q^2 r_+^2 (12376 r_+^6 + 12379 r_+^3 r_-^3 + 3 r_-^6) - 1092 c_6 e q r_+^5 r_-^3 (12376 r_+^3 + 3 r_-^3) \o]\nn\\
&=&546c_6 e q r_+^2[12376 r_+^6 c_6 e q+ 12376 r_+^3 r_-^3c_6 e q-2*12376 r_+^3r_-^3(r_+^3\o)
+3r_+^3 r_-^3c_6 e q + 3 r_-^6c_6 e q-2*3 r_-^6(r_+^3\o)].
\eea
Considering the superradiance condition $0<\o r_+^3<c_6 eq$ and $r_+>r_-$, it is easy to check the above term is positive.
It is easy to see that the terms outside the two square brackets in the difference $b'_{12}-b'_{11}$ are positive.
Thus  $b'_{12}>b'_{11}$ and we have the following sign relation
\bea
\text{sign}(b_{12})\geqslant\text{sign}(b_{11}).\label{sign1211}
\eea
For the differences $b'_{11}-b'_{10}$ and $b'_{10}-b'_{9}$, we have
\bea
&&b'_{11}-b'_{10}=\frac{99r_+^{5}}{4R_{11}R_{10}}\{[156 r_+^5 (6188 r_+^9 + 18156 r_+^6 r_-^3 + 17769 r_+^3 r_-^6 + 5801 r_-^9)(\mu^2-\o^2)]+ r_+^3 r_-^9 (-6792 + 11977 \l)\nn\\&&
+[546 c_6^2 e^2 q^2 r_+^2 (1768 r_+^6 + 1771 r_+^3 r_-^3 + 3 r_-^6) - 1092 c_6 e q r_+^5 r_-^3 (1768 r_+^3 + 3 r_-^3) \o] +370 r_-^{12} (6 + \l) \nn\\&& + 67 r_+^6 r_-^6 (66336 + 19619 \l) + 39 r_+^{12} (389188 + 153477 \l) + 3 r_+^9 r_-^3 (5236464 + 1786423 \l)\},\\
&&b'_{10}-b'_{9}=\frac{33r_+^{4}}{8R_{10}R_{9}}\{[10296 r_+^8 (1547 r_+^9 + 4386 r_+^6 r_-^3 + 4152 r_+^3 r_-^6 + 1313 r_-^9)(\mu^2-\o^2)]
\nn\\&&+[36036 c_6^2 e^2 q^2 r_+^5 (442 r_+^6 + 445 r_+^3 r_-^3 + 3 r_-^6) - 72072 c_6 e q r_+^8 r_-^3 (442 r_+^3 + 3 r_-^3) \o]
+r_+^3 r_-^{12} (105419 + 15796 \l) \nn\\&&+ 518 r_+^9 r_-^6 (86744 + 36503 \l) + 2 r_+^6 r_-^9 (201629 + 206456 \l) + 143 r_+^{15} (1176187 + 432746 \l)\nn\\&& + 2 r_+^{12} r_-^3 (76753559 + 26070946 \l)-88 r_-^{15}\}.
\eea
Using a similar way in proving $b'_{12}-b'_{11}$, we can get the following two sign relations
\bea
\text{sign}(b_{11})\geqslant\text{sign}(b_{10}),~\text{sign}(b_{10})\geqslant\text{sign}(b_{9}).\label{sign1109}
\eea

\begin{figure}[H]
    \centering
     \includegraphics[scale=0.4]{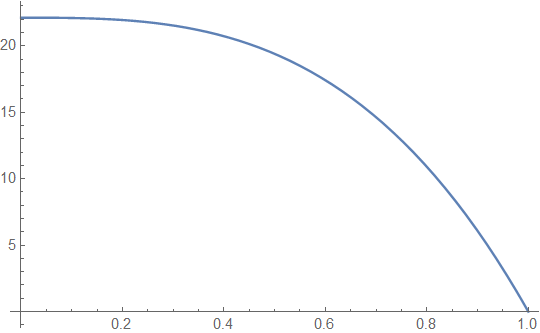}
    \caption{$\tilde{g}(x)/10^5$}
    \label{fig98}
\end{figure}

The differences $b'_{9}-b'_{8}$ can be obtained as follows,
\bea
&&b'_{9}-b'_{8}=\frac{3r_+^{5}}{8R_{9}R_{8}}\{[3432 r_+^9 (85085 r_+^9 + 227205 r_+^6 r_-^3 + 202790 r_+^3 r_-^6 +
   60642 r_-^9)(\mu^2-\o^2)]-14023 r_+r_-^{15}\nn\\&&+r_+[132 c_6^2 e^2 q^2 r_+^2 (2212210 r_+^9 + 2254863 r_+^6 r_-^3 +
    42486 r_+^3 r_-^6 - 167 r_-^9) -
 264 c_6 e q r_+^5 r_-^3 (2212210 r_+^6 \nn\\&&+ 42653 r_+^3 r_-^3 - 167 r_-^6) \o]+r_+^4 r_-^{12} (5227753 + 574000 \l) +
 143 r_+^{16} (15940879 + 5494082 \l)  \nn\\&&+ 55 r_+^{13} r_-^3 (28162775 + 11098882 \l) +
 2 r_+^{10} r_-^6 (169113407 + 150950723 \l)+ 2 r_+^7 r_-^9 (13964639 + 6361915 \l)\}.\label{b9-b8}
\eea
In the above equation, the term in the square bracket involving $(\mu^2-\o^2)$ is obviously positive. The term in the square bracket involving $\o$ can be treated as a linear function $g(\o)$ with negative slope
\bea
&&[132 c_6^2 e^2 q^2 r_+^2 (2212210 r_+^9 + 2254863 r_+^6 r_-^3 +42486 r_+^3 r_-^6 - 167 r_-^9) -
 264 c_6 e q r_+^5 r_-^3 (2212210 r_+^6 + 42653 r_+^3 r_-^3 - 167 r_-^6) \o]\nn\\
&& \equiv g(\o).
\eea
In the superradiant regime $0<\omega r_+^3<c_6 e q$, $g(\o)$ is greater than $g(c_6 e q/r_+^3)$. $g(c_6 e q/r_+^3)$ can be calculated as
\bea
&&132c_6^2 e^2 q^2 r_+^2(2212210 r_+^9 - 2169557 r_+^6 r_-^3 - 42820 r_+^3 r_-^6 + 167 r_-^9)\nn\\
&=&132c_6^2 e^2 q^2 r_+^{11}(2212210 - 2169557 x^3 - 42820 x^6 + 167 x^9)=132c_6^2 e^2 q^2 r_+^{11}\tilde{g}(x),~(0<x=r_-/r_+<1).
\eea
From Fig.\eqref{fig98}, we can see $\tilde{g}(x)$ is greater than $0$ when $0<x<1$.
So the term in the square bracket involving $\o$ in the difference $b'_{9}-b'_{8}$ is positive.
It is easy to see that the terms outside the two square brackets in the difference $b'_{9}-b'_{8}$ are positive.
Thus  $b'_{9}>b'_{8}$ and we have the following sign relation
\bea
\text{sign}(b_{9})\geqslant\text{sign}(b_{8}).\label{sign98}
\eea

The differences $b'_{i+1}-b'_{i},(i=2,...,7)$ are
\bea
&&b'_{8}-b'_{7}=\frac{9r_+^{7}}{4R_{8}R_{7}}\{[10296 r_+^9 (4862 r_+^9 + 11781 r_+^6 r_-^3 + 9539 r_+^3 r_-^6 + 2604 r_-^9)(\mu^2-\o^2)]\nn\\&&
+[3r_+(132 c_6^2 e^2 q^2 r_+^2 (126412 r_+^9 + 131859 r_+^6 r_-^3 +5358 r_+^3 r_-^6 - 89 r_-^9) -264 c_6 e q r_+^5 r_-^3 (126412 r_+^6 \nn\\&&+ 5447 r_+^3 r_-^3 - 89 r_-^6) \o)]+3r_+(10 r_+^3 r_-^{12} (78281 + 4778 \l) + 143 r_+^{15} (723902 + 235261 \l)-3791 r_-^{15} \nn\\&& +
 11 r_+^{12} r_-^3 (3056591 + 2117728 \l) + r_+^9 r_-^6 (-5097436 + 14755391 \l)+ 2 r_+^6 r_-^9 (2067013 + 519275 \l)) \},\\
&&b'_{7}-b'_{6}=\frac{3r_+^{9}}{4R_{7}R_{6}}\{[19656 r_+^9 (4862 r_+^9 + 10098 r_+^6 r_-^3 + 6971 r_+^3 r_-^6 + 1647 r_-^9)(\mu^2-\o^2)]\nn\\&&
+r_+[756 c_6^2 e^2 q^2 r_+^2 (126412 r_+^9 + 136461 r_+^6 r_-^3 +9609 r_+^3 r_-^6 - 440 r_-^9) -
 1512 c_6 e q r_+^5 r_-^3 (126412 r_+^6 \nn\\&&+ 10049 r_+^3 r_-^3 - 440 r_-^6) \o]+r_+(4 r_+^3 r_-^{12} (2647665 + 56197 \l) +
 91 r_+^{15} (5476878 + 1690829 \l)-124740 r_-^{15}\nn\\&& + 3 r_+^6 r_-^9 (17188506 + 2413601 \l) + 7 r_+^{12} r_-^3 (-9401814 + 12678373 \l) + r_+^9 r_-^6 (-131700942 + 69208211 \l))\},\\
&&b'_{6}-b'_{5}=\frac{21r_+^{11}}{4R_{6}R_{5}}\{[2184 r_+^9 (2431 r_+^9 + 3927 r_+^6 r_-^3 + 2058 r_+^3 r_-^6 + 386 r_-^9)(\mu^2-\o^2)]
\nn\\&&+[r_+(84 c_6^2 e^2 q^2 r_+^2 (63206 r_+^9 + 70915 r_+^6 r_-^3 + 6961 r_+^3 r_-^6 -748 r_-^9) -
 168 c_6 e q r_+^5 r_-^3 (63206 r_+^6+ 7709 r_+^3 r_-^3 \nn\\&&- 748 r_-^6) \o)]+r_+(r_+^3 r_-^{12} (1147509 - 8378 \l) +
 39 r_+^{15} (624339 + 184427 \l) + r_+^6 r_-^9 (5331078 + 406607 \l) \nn\\&&+ r_+^9 r_-^6 (-9604746 + 2889083 \l) +
 r_+^{12} r_-^3 (-15469083 + 2941951 \l)-38619 r_-^{15})\},\\
&&b'_{5}-b'_{4}=\frac{9r_+^{13}}{4R_{5}R_{4}}\{[168 r_+^9 (17017 r_+^9 + 17391 r_+^6 r_-^3 + 5028 r_+^3 r_-^6 + 650 r_-^9)(\mu^2-\o^2)]
\nn\\&&+[r_+(84 c_6^2 e^2 q^2 r_+^2 (34034 r_+^9 + 39049 r_+^6 r_-^3 + 4003 r_+^3 r_-^6 -1012 r_-^9) -
 168 c_6 e q r_+^5 r_-^3 (34034 r_+^6 \nn\\&&+ 5015 r_+^3 r_-^3 - 1012 r_-^6) \o)]+r_+(r_+^3 r_-^{12} (885447 - 27562 \l) +
 21 r_+^{15} (569799 + 162167 \l)-82215 r_-^{15}\nn\\&& + r_+^6 r_-^9 (4448970 + 173713 \l) + r_+^{12} r_-^3 (-13174155 + 529733 \l) + r_+^9 r_-^6 (-4043826 + 1001845 \l))\},\\
&&b'_{4}-b'_{3}=\frac{9r_+^{15}}{4R_{4}R_{3}}\{[12 r_+^9 (30940 r_+^9 + 8670 r_+^6 r_-^3 - 1875 r_+^3 r_-^6 + 11 r_-^9)(\mu^2-\o^2)]
\nn\\&&+[r_+(42 c_6^2 e^2 q^2 r_+^2 (8840 r_+^9 + 9739 r_+^6 r_-^3 + 433 r_+^3 r_-^6 -466 r_-^9) -
 84 c_6 e q r_+^5 r_-^3 (8840 r_+^6 + 899 r_+^3 r_-^3 - 466 r_-^6) \o)]\nn\\&&+r_+(r_+^3 r_-^{12} (84387 - 6515 \l) +
 r_+^6 r_-^9 (705807 + 12275 \l) - 5 r_+^{12} r_-^3 (425247 + 18865 \l) \nn\\&&+ r_+^9 r_-^6 (-101757 + 72236 \l) +
 3 r_+^{15} (487614 + 134647 \l)-25044 r_-^{15})\},
 \eea
 \bea
&&b'_{3}-b'_{2}=\frac{9r_+^{17}}{4R_{3}R_{2}}\{[4 r_+^9 (6188 r_+^9 - 3876 r_+^6 r_-^3 + 123 r_+^3 r_-^6 - 5 r_-^9)(\mu^2-\o^2)]
\nn\\&&+[r_+(14 c_6^2 e^2 q^2 r_+^2 (1768 r_+^9 + 1553 r_+^6 r_-^3 - 283 r_+^3 r_-^6 - 68 r_-^9) -
 28 c_6 e q r_+^5 r_-^3 (1768 r_+^6 - 215 r_+^3 r_-^3 - 68 r_-^6) \o)]\nn\\&&+r_+(r_+^6 r_-^9 (43365 - 197 \l) -
 3 r_+^3 r_-^{12} (2227 + 126 \l) + 3 r_+^9 r_-^6 (10257 + 1967 \l) - 3 r_+^{12} r_-^3 (53606 + 7013 \l)\nn\\&& +
 r_+^{15} (94182 + 25433 \l)-819 r_-^{15})\}.
\eea
Using a similar way in proving $b'_{9}-b'_{8}$, we can obtain the following sign relations
\bea
\text{sign}(b_{i+1})\geqslant\text{sign}(b_{i}),(i=2,...,7).\label{sign82}
\eea

\begin{figure}[H]
    \centering
     \includegraphics[scale=0.4]{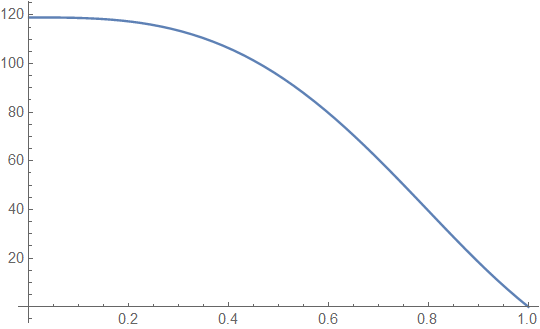}
    \caption{$g_1(x)$}
    \label{fig21}
\end{figure}

Finally, we consider the difference of $b_{2}'$ and $b_{1}'$,
\bea
&&b'_{2}-b'_{1}=\frac{9r_+^{19}}{8R_{2}R_{1}}\{[12 r_+^9 (119 r_+^9 - 204 r_+^6 r_-^3 + 105 r_+^3 r_-^6 - 20 r_-^9)(\mu^2-\o^2)]
\nn\\&&+[3 r_+ (119 r_+^6 - 85 r_+^3 r_-^3 + 20 r_-^6) (4 c_6^2 e^2 q^2 r_+^2 (r_+^3 + r_-^3) - 8 c_6 e q r_+^5 r_-^3 \o)]+
3 r_+ (119 r_+^6 - 85 r_+^3 r_-^3 \nn\\&&+ 20 r_-^6) (r_+^3 - r_-^3) (-6 r_+^3 r_-^3 - 9 r_-^6 + r_+^6 (15 + 4 \l))\}.
\eea
The two square brackets in the above difference can be rewritten as
\bea
&&[12 r_+^9 (119 r_+^9 - 204 r_+^6 r_-^3 + 105 r_+^3 r_-^6 - 20 r_-^9)(\mu^2-\o^2)]\nn\\
&=&12 r_+^{18}(119 - 204x^3 + 105x^6 - 20 x^9)(\mu^2-\o^2)\equiv 12 r_+^{18}(\mu^2-\o^2)g_1(x),\label{b2-b1-mu} \\
&&[3 r_+ (119 r_+^6 - 85 r_+^3 r_-^3 + 20 r_-^6) (4 c_6^2 e^2 q^2 r_+^2 (r_+^3 + r_-^3) - 8 c_6 e q r_+^5 r_-^3 \o)]\nn\\
&=&3 r_+ (119 r_+^6 - 85 r_+^3 r_-^3 + 20 r_-^6)4c_6 e q r_+^2(c_6 e q (r_+^3 + r_-^3)-2r_+^3 r_-^3 \o).\label{b2-b1-o}
\eea
From Fig.\eqref{fig21}, we can see $g_1(x)$ is greater than $0$ when $0<x<1$.
Given the bound state condition $\o^2<\mu^2$ and superradiance condition $0<\o r_+^3<c_6 eq$, the two square brackets \eqref{b2-b1-mu},\eqref{b2-b1-o} are positive. Then the difference $b_{2}'-b_{1}'$ is positive. We have the following sign relation
\bea
\text{sign}(b_{2})\geqslant\text{sign}(b_{1}).\label{sign21}
\eea

According to the results on signs and sign relations of coefficients \eqref{ineq1},\eqref{ineq2},\eqref{ineq3},\eqref{ineq4},\eqref{sign1413},\eqref{sign1312},\eqref{sign1211},\eqref{sign1109},
\eqref{sign98},\eqref{sign82},\eqref{sign21},
we conclude that the possible signs of the ordered coefficients $(b_{18},b_{17},...,b_{0})$ are of the following forms
\begin{align}
    (+,+,+,+,+,+,+,+,+,+,+,+,+,+,+,+,+,+,-),(+,+,+,+,+,+,+,+,+,+,+,+,+,+,+,+,+,-,-),\nn\\
    (+,+,+,+,+,+,+,+,+,+,+,+,+,+,+,+,-,-,-),(+,+,+,+,+,+,+,+,+,+,+,+,+,+,+,-,-,-,-),\nn\\
    (+,+,+,+,+,+,+,+,+,+,+,+,+,+,-,-,-,-,-),(+,+,+,+,+,+,+,+,+,+,+,+,+,-,-,-,-,-,-),\nn\\
    (+,+,+,+,+,+,+,+,+,+,+,+,-,-,-,-,-,-,-),(+,+,+,+,+,+,+,+,+,+,+,-,-,-,-,-,-,-,-),\nn\\
    (+,+,+,+,+,+,+,+,+,+,-,-,-,-,-,-,-,-,-),(+,+,+,+,+,+,+,+,+,-,-,-,-,-,-,-,-,-,-),\nn\\
    (+,+,+,+,+,+,+,+,-,-,-,-,-,-,-,-,-,-,-),(+,+,+,+,+,+,+,-,-,-,-,-,-,-,-,-,-,-,-),\nn\\
    (+,+,+,+,+,+,-,-,-,-,-,-,-,-,-,-,-,-,-),(+,+,+,+,+,-,-,-,-,-,-,-,-,-,-,-,-,-,-),\nn\\
    (+,+,+,+,-,-,-,-,-,-,-,-,-,-,-,-,-,-,-).\label{result}
\end{align}
Note that the plus signs are always on the left of the minus signs.

Let's recall the numerator $\tilde{C}(z)$ of the derivative of the effective potential in eq.\eqref{derivative of potential}. It is a polynomial of $z$ with real coefficients $(b_{18},b_{17},...,b_{0})$. According to Descartes' rule of signs and the results in \eqref{result}, we obtain that the sign change in the sequence of the coefficients $(b_{18},b_{17},...,b_{0})$ is always 1 and therefore there is at most one positive real root for equation $\tilde{C}(z)=0$. That means there is at most one extreme for the effective potential outside the event horizon $r_+$. Based on the analysis of the asymptotic behaviors of the effective potential,
we have already known that there is one maximum outside the event horizon $r_+$ for the effective potential. So we conclude that there is no potential well for the effective potential outside the horizon $r_+$ and no black hole bomb exists for the six-dimensional non-extremal RN black holes.
\section{Summary}
In this work, we analytically study the superradiant stability of the six-dimensional non-extremal RN black hole under charged massive scalar perturbation. The analytic method is based on Descartes' rule of signs. The equation of motion of the minimally coupled scalar perturbation is Klein-Gordon equation. We decompose it into angular and radial parts. From the radial equation of motion, we obtain the effective potential experienced by the scalar. The necessary condition for the superradiant instability of
the system is there exists a potential well for the effective potential $V$.
Based on the asymptotic analysis of the effective potential, we find that there is one maximum outside the outer horizon for the effective potential. We then discuss the real roots of the derivative of the effective potential $V'$ with the analytic method based on Descartes' rule of signs. It is found that the sign change of the numerator of  $V'$ is always 1 and there is no potential well for the superradiant modes outside the outer horizon. No black hole bomb exists for the system consisting of six-dimensional non-extremal RN black hole and charged massive scalar perturbation, and the system is superradiantly stable.

According to the result of this work and the previous results on superradiant stability of higher dimensional extremal and non-extremal RN black holes under charged massive scalar perturbation, it will be interesting to prove that no black hole bomb exists for arbitrary $D$-dimensional non-extremal RN black hole under charged massive scalar perturbation and the system is superradiantly stable.

\begin{acknowledgements}
This work is partially supported by Guangdong Major Project of Basic and Applied Basic Research (No.2020B0301030008), Science and Technology Program of Guangzhou (No.2019050001) and Natural Science Foundation of Guangdong Province (No.2020A1515010388).
\end{acknowledgements}

\end{document}